\documentclass[aps,prl,reprint]{revtex4-1}
\usepackage{blindtext}

\usepackage{amsmath,amssymb,amsthm,amscd,graphicx}
\usepackage{psfrag}
\usepackage[utf8]{inputenc}
\usepackage{graphicx}
\usepackage{color}
\usepackage{hyperref}
\hypersetup{colorlinks,allcolors=blue}
\usepackage{filecontents}
\usepackage{color}

\usepackage[T1]{fontenc}

\usepackage[utf8]{inputenc}
\usepackage{graphicx}
\usepackage{amsmath,amsthm,amssymb,physics,mathtools}
\usepackage{tikz,tikz-cd, dynkin-diagrams}
\usepackage{epigraph}
\usepackage{hhline}
\usepackage{caption}
\usepackage{subcaption}
\usepackage{ytableau}
\usepackage{natbib}

\theoremstyle{definition}

\newcommand{\CZ}{{\cal Z}}

\newcommand{\bea}{\begin{eqnarray}}
\newcommand{\eea}{\end{eqnarray}}

\def\l{\ell}
\def\a{\boldsymbol{\alpha}}

\def\b{\boldsymbol{\beta}}
\def\e{\boldsymbol{\eta}}

\def\s{\boldsymbol{\sigma}}

\def\l{\boldsymbol{\lambda}}

\def\n{\vb n}
\def\m{\vb m}

\def\v{\v}
\def\I{\sqrt{-1}}

\def\Z{\mathbb{Z}}
\def\12{\frac{1}{2}}
\newcommand{\be}{\begin{equation}}
\newcommand{\ee}{\end{equation}}
\newcommand{\ba}{\begin{aligned}}
	\newcommand{\ben}{\begin{eqnarray}\displaystyle}
	\newcommand{\een}{\end{eqnarray}}

\begin{document}

		\title{Instantons to the people: the power of one-form symmetries}
		\author{Giulio Bonelli, Fran Globlek, Alessandro Tanzini}
		\affiliation{International School of Advanced Studies (SISSA), via Bonomea 265, 34136 Trieste, Italy and INFN, Sezione di Trieste}
		\affiliation{Institute for Geometry and Physics, IGAP, via Beirut 2, 34136 Trieste, Italy}
		
		\date{\today}
		
		\begin{abstract}
			Abstract: We show that the non-perturbative dynamics of $\mathcal{N}=2$ super Yang-Mills theories in a self-dual $\Omega$-background and with an arbitrary simple gauge group is fully determined by studying renormalization group equations of vevs of surface operators generating one-form symmetries. The corresponding system of equations is a {\it non-autonomous} Toda chain, the time being the RG scale.  We obtain new recurrence relations which provide a systematic algorithm computing multi-instanton corrections from the tree-level one-loop prepotential as the asymptotic boundary condition of the RGE. We exemplify by computing the $E_6$ and $G_2$ cases up to two-instantons.
		\end{abstract}

		%\begin{document}
		\maketitle
		
		%\flushbottom
		
		In an ideal world the non-perturbative structure of gauge theories should be computed by quantum equations of motion determined by a symmetry principle. The presence of extended operators generating higher form symmetries in quantum field theory
		is a powerful tool to concretely realise such a programme. A perturbative analysis in a weakly coupled regime, if any, would supply appropriate asymptotic conditions.
		In this letter we present a class of theories where the full non-perturbative result is fixed in such a framework. These are $\mathcal {N}=2$ super Yang-Mills theories in four dimensional self-dual $\Omega$-background, which enjoy a one-form symmetry generated by surface operators \cite{Gaiotto:2014kfa}. 
		We show that the renormalization group equation obeyed by the vacuum expectation value of such surface operators provides a recursion relation which fully determines, from the perturbative one-loop
		prepotential, all instanton contributions on the self-dual $\Omega$-background or, equivalently, the all-genus topological string amplitudes on the relevant geometric background. 
		Actually, partition functions with surface operators display a very clear resurgent structure led by the summation over the magnetic fluxes \footnote{For an introduction to resurgence in QFT, see for example M. Marino, {\it Instantons and Large N}, (2015) Cambridge University Press.}.
		
		The system of equations we study is a {\it non-autonomous}
		twisted affine Toda chain of type $(\hat G )^\vee$, where $(\hat G)^\vee$ is the Langlands dual of the untwisted affine Kac-Moody algebra $\hat G$. Each node of the corresponding affine Dynkin diagram 
		defines a surface operator, the associated $\tau$-function being its vacuum expectation value. 
		The time flow corresponds in the gauge theory to the renormalization group.
		The resulting recurrence relations constitute a new effective algorithm to determine instanton contributions for all classical groups $G$.
		Let us remark that the $\tau$-functions we obtain provide the general solution at the canonical rays for the Jimbo-Miwa-Ueno isomonodromic deformation problem 
		\cite{Jimbo:1981zz,Jimbo:1982zz}
		on the sphere with two-irregular punctures
		for all classical groups, which to the best of our knowledge was not known in the previous literature.
		The recursion relations we obtain are different from the blow-up equations of \cite{Nakajima:2003pg} further elaborated in \cite{Kim:2019uqw}.
		Indeed the latter necessarily involve the knowledge of the partition function in different $\Omega$-backgrounds.
		This makes the recursion relations (and the results) coming from blow-up equations more involved and difficult to handle.
		However, we expect a relation between the two approaches to follow from blow-up relations in presence of surface defects.
		Indeed, the isomonodromic $\tau$-function for the sphere with four regular punctures was obtained in a similar way from $SU(2)$ gauge theory with $N_f=4$
		in \cite{Nekrasov:2020qcq}.
		In this letter we summarise our results and refer to a subsequent longer paper for a fully detailed discussion.
		
		The $\tau$-functions are labeled by the simple roots of the affinization of the Lie algebra of the gauge group $\alpha\in \hat\Delta$, namely
		$\{\tau_\alpha\}_{\alpha\in \hat\Delta}$,
		and satisfy the equations
		%\begin{equation}\label{hirotatausystem} D^2([\tau_{\b}]^{{\b^\vee\cdot\b^\vee\over 2}})={\b^\vee\cdot\b^\vee\over 2}\,t^{1/h^\vee} \prod\limits_{\beta\in\hat\Delta,\b\neq\a}\left[ \tau_{\boldsymbol\alpha} \right]^{-\a^\vee\cdot\b^\vee}
		%\end{equation}
		\begin{equation}\label{hirotatausystem} D^2(\tau_{\b})=-{\b^\vee\cdot\b^\vee\over 2}\,t^{1/h^\vee} \prod\limits_{\beta\in\hat\Delta,\b\neq\a}\left[ \tau_{\boldsymbol\alpha} \right]^{-\a\cdot\b^\vee}
		\end{equation}
		where $t:=\left(\Lambda/\epsilon\right)^{2 h^\vee}$ and the
		logarithmic Hirota derivative is given by
		$D^2(f)=f\partial_{\log t}^2f - (\partial_{\log{t}}f)^2.$
		Given a simple root $\alpha$, its coroot is as usual given by $\alpha^\vee=2\alpha/(\alpha,\alpha)$, where $(\cdot,\cdot)$ is the scalar product defined by the affine Cartan matrix.
		Eq. \eqref{hirotatausystem} is the de-autonomization of the $\tau$-form of the standard Toda integrable system \cite{Gorsky:1995zq, Martinec:1995by} governing the classical Seiberg-Witten (SW) theory 
		\cite{Seiberg:1994rs}.
		The de-autonomization is induced by coupling the theory to a self-dual $\Omega$-background $(\epsilon_1,\epsilon_2)=(\epsilon,-\epsilon)$ \cite{Bonelli:2016qwg}.
		In the autonomous limit $\epsilon\to 0$, $\tau$-functions reduce to $\theta$-functions on the classical SW curve \cite{Bonelli:2019boe}, which were used to provide recursion relations on the coefficients of the SW prepotential in \cite{Edelstein:1998sp}.
		The gauge theory interpretation of these $\tau$-functions is the v.e.v. of surface operators associated to the corresponding decomposition of the Lie algebra representation under which these are charged. 
		We expect these equations and their generalizations to describe chiral ring relations in presence of a surface operator, which deserve further investigation.
		Higher chiral observables should generate the flows of the full {\it non-autonomous} Toda hierarchy.
		The actual form of equations \eqref{hirotatausystem} depends on the Dynkin diagram. For the classical groups $A$, $B$ and $D$ these reduce to bilinear 
		equations  which we solve via general recursion relations.
		For $C$, $E$, $F$ and $G$ the resulting equations are of higher order and we study them case by case. 
		The symmetries of the equations are given by the center of the group $G$, namely
		\begin{center}
			\begin{tabular}{c|c|c|c|c|c|c|c|c} 
				$\mathfrak g$ & $A_n$ & $B_n$ & $C_n$ & $D_{2n}$ & $D_{2n+1}$ & $E_{n}$ & $F_4$ & $G_2$ \\ 
				\hline 
				$Z(G)$ & $\Z_{n+1}$ & $\Z_2$ & $\Z_2$ & $\Z_2\times \Z_2$ & $\Z_4$ & $\Z_{9-n}$ & $1$ & $1$ \\  
			\end{tabular} 
		\end{center}
		
		Moreover, the center is isomorphic to the coset of the affine coweight lattice by the
		affine coroot lattice, and coincides with the automorphism group of the affine Dynkin diagram. By a remark in \footnote{ Bourbaki [Lie gps Ch. VIII \S 7]}, the coweights, and by extension the lattice cosets,  
		corresponding to these nodes are the miniscule coweights, a representation of $\mathfrak g$ being miniscule if all its weights form a single Weyl-orbit.
		This remark will be crucial while solving the $\tau$-system.
		%, and this is the analogue for $\mathfrak g^\vee$ {\bf [???]}.
		
		The $\tau$-functions corresponding to the affine nodes, that is the ones
		which can be removed from the Dynkin diagram leaving behind that of an irreducible simple Lie algebra, 
		play a special r\^ole. Indeed, these are related to simple surface operators associated to elements of the center $Z(G)$, and are bounded by 
		fractional 't Hooft lines. Such surface operators are the generators of the one-form symmetry of the 
		corresponding gauge theory, \cite{Gaiotto:2014kfa}. Since their magnetic charge is defined modulo the magnetic root lattice, a natural Ansatz for their expectation value is 
		\begin{equation}
		\label{Kiev}
		\tau_{\a_{\text{aff}}}\left(\s,\e|\kappa_{\mathfrak g}t\right) = \sum\limits_{\n\in Q_{\text{aff}}^\vee} 
		e^{2\pi\I\e\cdot\n}
		t^{\12(\s+\n)^2}B(\s+\n|t)
		\end{equation}
		where $B(\s|t)=
		%\mathcal{Z}
		B_0(\s) \sum_{i\ge 0}t^i Z_i(\s)$ 
		with $Z_0(\s)\equiv 1$ 
		and  
		$Q_{\text{aff}}^\vee=\l_{\text{aff}}^\vee + Q^\vee$, $Q^\vee$ being the co-root lattice and
		$(\l_{\text{aff}}^\vee,\alpha)=\delta_{\alpha_{\text{aff}},\alpha}$ for any simple root $\alpha$.
		The constant 
		%	\begin{equation}
		$
		\kappa_{\mathfrak g}= (-n_{\mathfrak g})^{r_{\mathfrak g,s}}
		$,
		%	\end{equation} 
		where $n_{\mathfrak g}$ is the ratio of the squares of  long vs. short roots and $r_{\mathfrak g,s}$ is the number of short simple roots. For simply laced, all roots are long and $\kappa_{\mathfrak g}=1$.
		
		We will now show how the term $t^{\frac{1}{2}\s^2}B(\s|t)$ in \eqref{Kiev} is the full Nekrasov partition function in the self-dual $\Omega$-background upon the identification $\s={\bf a}/\epsilon$, where ${\bf a}$ is the Cartan parameter.
		In the $A_n$ case, \eqref{Kiev} is known as the Kiev Ansatz. In the $A_1$ case, it was used to give the general solution of Painlev\'e III$_3$ equation in \cite{Its:2014lga} and further analysed in \cite{Mironov:2017lgl}.
		
		Let us remark that the $\tau$-function \eqref{Kiev} displays a clear resurgent structure, with “instantons” given by the magnetic fluxes in the lattice summed with “resurgent” coefficients $B(\s | t)$ and trans-series parameter $e^{2\pi\I\e}$, see \cite{Dunne:2019aqp} for a similar analysis in the Painlev\'e III$_3$ case.
		
		The Ansatz \eqref{Kiev} is consistent with equations \eqref{hirotatausystem}. Indeed, after eliminating the 
		$\tau$-functions associated to the 
		non-affine nodes, the resulting equation is bilinear and therefore the Ansatz \eqref{Kiev}
		reduces to a set of recursion relations for the coefficients $Z_i(\s)$.
		The variables $\e$ and $\s$ are the integration constants of the second order differential equations \eqref{hirotatausystem} and correspond to the initial position and velocity of the 
		de-autonomized Toda particle.

		Let us set more precisely the boundary conditions which we impose to the solutions
		of equations \eqref{hirotatausystem}.
		We consider the asymptotic behaviour of the solutions at $t\to0$ and $\s\to \infty$ as 
		\be \label{asy}
		\log (B_0)\sim  
		-\frac{1}{4} 
		\sum_{\vb r \in R}  (\vb r \cdot \s)^2  \log\left(\vb r \cdot\s\right)^2 
		\ee
		up to quadratic and $\log$-terms \footnote{Notice that if one chooses 
			\unexpanded{$\log (B_0)\sim\sum_{\vb r \in R} c_{n,m}(\vb r \cdot \s)^{2n}\log\left((\vb r \cdot\s)^2\right)^m $}, then the equation itself dictates that $(n,m)=(1,1)$ and $(2,0)$}.	
		We will show that the solution of \eqref{hirotatausystem} which satisfies the above asymptotic condition is such that 
		\vspace*{-0.0cm}
		\be\label{1loop}
		B_{0}(\s)=\CZ_{1-loop}(\s)\equiv \prod_{\vb r \in R}\frac{1}{G(1+\vb r \cdot\s)}
		\ee
		%\vspace*{-0.1cm}
		where $G(z)$ is the Barnes' G-function and $R$ is the adjoint representation of the group $G$. 
		The expansion of the above function matches the one-loop
		gauge theory result upon the appropriate 
		identification of the log-branch. This reads, in the gauge theory variables, as
		${\rm ln}\left[\sqrt{-1} \vb r \cdot\vb a /\Lambda\right]\in {\mathbb R}$ and matches the canonical Stokes rays 
		obtained in \cite{Guest:2012yg}.
		
		Let us first focus on the $A_n$ case whose affine Dynkin diagram is
		
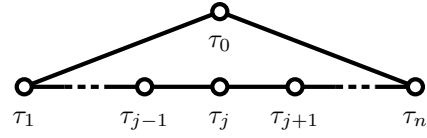
\vspace*{-0.3cm}
		\begin{center}
			\begin{dynkinDiagram}[extended,edge length=1cm, indefinite edge/.style={ultra thick,densely dashed}, edge/.style={ultra thick},o/.style={ultra thick,fill=white,draw=black}, root radius=.1cm,root/.style={ultra thick,fill=white,draw=black}]A{o.ooo.o}
				\node[below=.2cm] at (root 0) {$\tau_0$};
				\node[below=.2cm] at (root 1) {$\tau_1$};
				\node[below=.2cm] at (root 2) {$\tau_{j-1}$};
				\node[below=.2cm] at (root 3) {$\tau_{j}$};
				\node[below=.2cm] at (root 4) {$\tau_{j+1}$};
				\node[below=.2cm] at (root 5) {$\tau_{n}$};
			\end{dynkinDiagram}
		\end{center}
		\vspace*{-0.2cm}
		The root lattice is $Q=\{\sum\limits_{i=1}^{n+1}c_i e_i |\sum\limits_{i=1}^{n+1}c_i=0\}$, and all the fundamental weights are miniscule, namely
		$$
		\l_i = \frac{1}{n+1}(1^{i},0^{n+1-i})-\frac{i}{n+1}(1^{n+1}) \, ,
		$$
		where $(1^p,0^{n+1-p})$ stands for a vector whose first $p$ entries are $1$ and the remaining entries vanish.
		We label the $\tau$-functions as $\tau_{\a_j}\equiv\tau_j$. The 
		$\tau$-system is given by the closed chain of differential equations 
		\vspace*{-0.1cm}
		\begin{equation}\label{an}
		D^2(\tau_j)=-t^{\frac{1}{n+1}}\tau_{j-1}\tau_{j+1},
		\end{equation}
		with $\tau_j=\tau_{n+1+j}$.
		Since all the nodes in this case are affine we can use the Kiev Ansatz \eqref{Kiev}. Then, all the $\tau$-functions are determined by $\tau_0$ as $\tau_j(\s|t)=\tau_0(\s+\l_j|t)$. 
		It is therefore enough to solve the single equation  
		%	 With regards to the weights, we can check that
		%	$
		%	\12 \l_{j-1}^2+\12 \l_{j+1}^2 = -\frac{1}{n+1}+\l_j^2
		%	$
		%	so the exponents match. Since $\l_i\cdot \n = (e_1+...+e_i)\cdot\n$ for any $\n\in Q$, we can rewrite the equation as $D^2(\tau_0(\s+e_1+...+e_j))=\tau_0(\s+e_1+...+e_{j-1})\tau_0(\s+e_1+...+e_{j+1})$. Here, we assume that, if $\boldsymbol{\gamma}=\gamma \cdot (1^{n+1})$ for any constant $\gamma$, $\tau_0(\s+\boldsymbol{\gamma}|t)\propto \tau_0(\s|t)$, since $\s\cdot\boldsymbol{\gamma}=0$ and $Q\perp\boldsymbol{\gamma}$ by definition. Further, due to the cyclic symmetry, it is enough to solve, say 
		\vspace*{-0.1cm}
		\be\label{pm}
		D^2(\tau_0(\s))=-\tau_0(\s\pm e_1) \, .
		\ee
		Here and in the following we use the notation $f(y\pm x)\equiv f(y+x)f(y-x)$. The Ansatz \eqref{Kiev} for $\tau_0$ reads
		$$\resizebox{1\hsize}{!}{$\tau_0(\s,\e|t) = \sum\limits_{\n\in Q,\,i\geq 0}e^{2\pi\sqrt{-1}\n\cdot\e}
			t^{\12(\s+\n)^2+i}B_0(\s+\n)Z_i(\s+\n)$}
		$$
		and by inserting it into \eqref{pm} one gets after some simplifications
		\begin{align}
		&\sum_{\substack{\n_1,\n_2\in Q\\i_1,i_2\geq 0}}e^{2\pi\sqrt{-1}(\n_1+\n_2)\cdot\e}
		t^{\12 \n_1^2+\12 \n_2^2+i_1+i_2+\s\cdot(\n_1+\n_2)}\notag \\
		&\times\left(\12 \n_1^2-\12 \n_2^2+i_1-i_2+\s\cdot(\n_1-\n_2)\right)^2 \notag \\
		&\times B_0(\s+\n_1)B_0(\s+\n_2)Z_{i_1}(\s+\n_1)Z_{i_2}(\s+\n_2) \notag \\
		&=-\sum_{\substack{\m_1,\m_2\in Q\\j_1,j_2\geq 0}}t^{1+\12 \m_1^2+\12\m_2^2+e_1\cdot(\m_1-\m_2)+j_1+j_2+\s\cdot(\m_1+\m_2)}\times \notag \\
		&e^{2\pi\sqrt{-1}(\m_1+\m_2)\cdot\e}
		B_0(\s+\m_1+e_1)B_0(\s+\m_2-e_1)\times \notag \\
		&Z_{j_1}(\s+\m_1+e_1)Z_{j_2}(\s+\m_2-e_1)\label{strafu}
		\end{align}
		Now we simply equate the exponents. To fix $B_0(\s)$, we look at the lowest order in $t$. This produces a quadratic constraint and $n+1$ linear constraints on the root lattice variables $(\n_1,\n_2)$ and $(\m_1,\m_2)$.
		Let us fix $p,q\in\{0,...n+1\}$, $p\neq q$. Up to Weyl reflections, the only solution to the above mentioned constraints is given by
		$\n_1=e_p-e_q$, $\n_2=0$ and $\m_1=e_p-e_1$, $\m_2=-e_q+e_1$, leading to 
		\bea\label{An1loop}
		\left(1+(e_p-e_q)\cdot\s\right)^2 B_0(\s+e_p-e_q)B_0(\s) =\notag\\-B_0(\s+e_p)B_0(\s-e_q)  \, .
		\eea
		This is solved by \eqref{1loop} up to a function periodic on the root lattice, which is set to one by the asymptotic condition \eqref{asy}. 
		%	We stop here to comment on a subtlety. Notice that $F_0(\s)=F_{1-loop}(\s)\prod_k g_k(\sigma_k)$ also solves \eqref{An1loop}. However, recall that we have replaced $\l_1$ by $e_1$, because we assumed $\tau_0(\s+\boldsymbol{\gamma}|t)\propto \tau_0(\s|t)$. This translates directly to $F_0(\s+\gamma)=F_0(\s)$ for any such $\gamma$  
		The higher order terms in \eqref{strafu} provide the recursion relations
		\vspace*{-0.3cm}
		\bea
		k^2 Z_k(\s)=-\sum\limits_{\substack{\n^2+j_1+j_2=k\notag\\ \n\in e_1+ Q,\,j_{1,2}<k}}\frac{B_0(\s\pm\n)}{B_0(\s)^2}\times\\Z_{j_2}(\s-\n)Z_{j_1}(\s+\n)
		+\sum\limits_{\substack{\n^2+i_1+i_2=k\notag\\ \n\in Q,\, i_{1,2}<k}}\left(i_1-i_2+2\n\cdot\s\right)^2 \\ \times\frac{B_0(\s\pm\n)}{B_0(\s)^2}Z_{i_1}(\s+\n)Z_{i_2}(\s-\n) \, , \notag
		%	\label{Arec}
		%	k^2 Z_k(\s)=\sum\limits_{\substack{\n^2+j_1+j_2=k\\ \n\in e_1+ Q\,j_{1,2}<k}}Z_{j_1}(\s+\n)Z_{j_2}(\s-n)\frac{F_0(\s+\n)F_0(\s-\n)}{F_0(\s)^2}
		%	-\sum\limits_{\substack{\n^2+i_1+i_2=k\\\n\in Q,\, i_{1,2}<k}}Z_{j_1}(\s+\n)Z_{j_2}(\s-n)\left(i_1-i_2+2\n\cdot\s\right)^2\frac{F_0(\s+\n)F_0(\s-\n)}{F_0(\s)^2}
		%	\label{Arec}
		%	k^2 Z_k(\s)=\sum\limits_{\substack{\n^2+j_1+j_2=k\\ \n\in e_1+ Q,\,j_{1,2}<k}}\frac{Z_{j_1}(\s+\n)Z_{j_2}(\s-\n)}{\prod\limits_{\substack{\a\in R, n\geq 1\\ \a\cdot\n=n}}(\a\cdot\s)^{2n}\prod\limits_{k=1}^{n-1}((\a\cdot\s)^2-k^2)^{2n-2k}}
		%	-\sum\limits_{\substack{\n^2+i_1+i_2=k\\\n\in Q,\, i_{1,2}<k}}\frac{\left(i_1-i_2+2\n\cdot\s\right)^2 Z_{i_1}(\s+\n)Z_{i_2}(\s-\n)}{\prod\limits_{\substack{\a\in R, n\geq 1\\ \a\cdot\n=n}}(\a\cdot\s)^{2n}\prod\limits_{k=1}^{n-1}((\a\cdot\s)^2-k^2)^{2n-2k}}
		\eea
		where $B_0(\s)$ is given by \eqref{1loop}.
		%with $Z_0(\s)\equiv 1$, and recall that $R$ is the root system, not just positive roots. 
		For $k=1$ we easily obtain
		\begin{equation}
		Z_1(\s) = -\sum\limits_{i=1}^{n+1}\frac{B_0(\s\pm e_i)}{B_0(\s)^2}=(-1)^{n+1}\sum\limits_{i=1}^{n+1}\frac{1}{\prod_{j\neq i}(\sigma_i-\sigma_j)^2}\notag
		\end{equation}
		and, upon abbreviating $\sigma_{ij}=\sigma_i-\sigma_j$, the next term 
		\vspace*{-0.2cm}
		\bea
		Z_2(\s)&=& -\frac{1}{4}\sum\limits_{i=1}^{n+1}\frac{B_0(\s\pm e_i)}{B_0(\s)^2}
		[Z_1(\s+ e_i)+Z_1(\s- e_i)]\notag\\&+&
		\sum\limits_{i<j}^{n+1}(\sigma_i-\sigma_j)^2\frac{B_0(\s\pm(e_i-e_j))}{B_0(\s)^2}\notag
		\eea
		\vspace*{-0.2cm}
		
		\noindent The above coincide with one and two instanton contributions to the $SU(n+1)$ Nekrasov partition function as computed from supersymmetric localization
		\cite{Nekrasov:2003af, Nekrasov:2003rj}. 
		Let us remark that the use of the $\tau$-system \eqref{an} provides a completely independent tool to compute all instanton corrections just starting from the asymptotic behaviour \eqref{asy}. This procedure
		extends to all classical groups.
		%	\begin{dmath}
		%	Z_2(\boldsymbol{\sigma}) =
		%	\frac{1}{4}\sum_i \frac{1}{\prod_{j\neq i}\sigma_{ij}^2}\sum_k \frac{1}{\prod_{l\neq k}(\sigma_{kl}-1)^2}
		%	+\frac{1}{4}\sum_i \frac{1}{\prod_{j\neq i}\sigma_{ij}^2}\sum_k \frac{1}{\prod_{l\neq k}(\sigma_{kl}+1)^2}-\sum_{i<j}\frac{1}{(\sigma_{ij}+1)^2(\sigma_{ij}-1)^2 \prod_{i \neq k\neq j}\sigma_{ik}^2\sigma_{jk}^2}
		%	= \frac{1}{4}\sum_i \frac{1}{\prod_{k\neq i}\sigma_{ki}^2 \cdot \prod_{k\neq i}(\sigma_{ki}-1)^2}
		%	+\frac{1}{4}\sum_i \frac{1}{\prod_{k\neq i}\sigma_{ki}^2 \cdot \prod_{k\neq i}(\sigma_{ki}+1)^2}
		%	+\sum_{i<j}\frac{\sigma_{ij}^2(\sigma_{ij}^2-0^2)}{(\sigma_{ij}^2-1^2)(\sigma_{ij}^2-1^2)}\frac{1}{\prod_{k\neq i}\sigma_{ki}^2\cdot \prod_{k\neq j}\sigma_{kj}^2}
		%	\end{dmath}
		%	where in the second line we've cancelled the off-diagonal terms in the double summations and rewritten the result to make it easy to see that it is exactly a sum of pure $A_n$ Nekrasov functions for Young diagrams $[1][1]$, $[2]$ and $[11]$ corresponding to the 2-instanton contribution \cite{Mironov2009a} in the case $\epsilon_1=-\epsilon_2=1$, for which the $0^2=(\ea+\eb)^2$ stands for in the last sum.
		
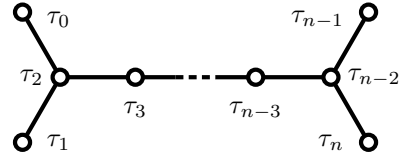
\vspace*{-0.1cm}
		\begin{center}
			\begin{dynkinDiagram}[extended,reverse arrows, edge length=1cm, indefinite edge/.style={ultra thick,densely dashed}, edge/.style={ultra thick},o/.style={ultra thick,fill=white,draw=black}, root radius=.1cm,root/.style={ultra thick,fill=white,draw=black}, arrow width = 0.4cm, arrow style={length=5mm, width=5mm,line width = 1pt}]D{ooo.oooo}
				\node[below=.1cm,right=.2cm] at (root 0) {$\tau_0$};
				\node[right=.2cm] at (root 1) {$\tau_1$};
				\node[left=.1cm] at (root 2) {$\tau_2$};
				\node[below=.2cm] at (root 3) {$\tau_{3}$};
				\node[below=.2cm] at (root 4) {$\tau_{n-3}$};
				\node[right=.1cm] at (root 5) {$\tau_{n-2}$};
				\node[below=.1cm,left=.2cm] at (root 6) {$\tau_{n-1}$};
				\node[left=.2cm] at (root 7) {$\tau_{n}$};
			\end{dynkinDiagram}
		\end{center}
		$D_n$ is a simply laced root system, with the checkerboard lattice $Q=Q^\vee =\{\sum_{i=1}^n c_i e_i | \sum_{i=1}^n c_i\in 2\Z \}$. We consider $n>4$. It has four miniscule weights, $\l_0=(0^n)$, $\l_1=(1,0^{n-1})$, $\l_{n-1}=((\12)^{n-1},-\12)$, $\l_{n}=((\12)^{n-1},+\12)$. These correspond to the "legs" of the affine diagram. Whichever rank we consider, we always have the consistency conditions
		\vspace*{-0.1cm}
		\begin{gather}\label{Dtau}
		D^2(\tau_0)=D^2(\tau_1),\quad  D^2(\tau_{n-1})=D^2(\tau_n)
		\end{gather}
		which are also equal if $n=4$, due to the enhanced symmetry of $D_4$. 
		
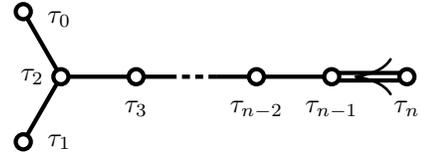
\begin{center}
			\begin{dynkinDiagram}[extended,reverse arrows, edge length=1cm, indefinite edge/.style={ultra thick,densely dashed}, edge/.style={ultra thick},o/.style={ultra thick,fill=white,draw=black}, root radius=.1cm,root/.style={ultra thick,fill=white,draw=black}, arrow width = 0.4cm, arrow style={length=5mm, width=5mm,line width = 1pt}]B{ooo.ooo}
				\node[below=.1cm,right=.2cm] at (root 0) {$\tau_0$};
				\node[right=.2cm] at (root 1) {$\tau_1$};
				\node[left=.1cm] at (root 2) {$\tau_2$};
				\node[below=.2cm] at (root 3) {$\tau_{3}$};
				\node[below=.2cm] at (root 4) {$\tau_{n-2}$};
				\node[below=.2cm] at (root 5) {$\tau_{n-1}$};
				\node[below=.2cm] at (root 6) {$\tau_{n}$};
			\end{dynkinDiagram}
		\end{center}
		
		$B_n$ is non-simply laced. 
		The coroot lattice is the checkerboard lattice $Q^\vee=\{\sum_{i=1}^n c_i e_i | \sum_{i=1}^n c_i\in 2\Z \}$, and the two miniscule weights are $\l_0^\vee=(0^n)$ and $\l_1^\vee=(1,0^{n-1})$, corresponding to the "antennae" of the diagram. The $\tau$-system coincides with that of $D_{n+1}$, with the modification that (i) there is no $\tau_{n+1}$ and (ii) that
		\begin{equation*}
		D^2(\tau_{n-1})=-2t^{\frac{1}{2n-1}}\tau_{n-2}\tau_{n},\quad
		D^2(\tau_n)=-t^{\frac{1}{2n-1}}\tau_{n-1}^2.
		\end{equation*}
		For $n\geq3$, the analysis proceeds as for $D_n$ except we may only use the left antennae and consider the first equation in \eqref{Dtau}.
		%\begin{equation}\label{Dtau}
		%D^2(\tau_0)=D^2(\tau_1)
		%\end{equation}
		Therefore, we have a unified approach for both $D_n$ and $B_n$. Explicitly, inserting \eqref{Kiev} and $\tau_1(\s|t)=\tau_0(\s+\l_1|t)$ into the first of \eqref{Dtau} we get after some simplification a formula analogous to
		\eqref{strafu} leading to quadratic and linear constraints on the lattice labels. 
		%	\begin{align}
		%	\sum_{\substack{\n_1,\n_2\in Q^\vee\\i_1,i_2\geq 0}}&t^{\12 \n_1^2+\12 \n_2^2+i_1+i_2+\s\cdot(\n_1+\n_2)}\left(\12 \n_1^2-\12 \n_2^2+i_1-i_2+\s\cdot(\n_1-\n_2)\right)^2
		%	\\
		%	&F_0(\s+\n_1)F_0(\s+\n_2)Z_{i_1}(\s+\n_1)Z_{i_2}(\s+\n_2)
		%	\\
		%	=\sum_{\substack{\m_1,\m_2\in Q\\j_1,j_2\geq 0}}&t^{1+\12 \m_1^2+\12 \m_2^2+\l_1\cdot(\m_1+\m_2)+j_1+j_2+\s\cdot(\m_1+\m_2+2\l_1)}
		%	\\
		%	&\left(\12 \m_1^2-\12 \m_2^2+j_1-j_2+(\s+\l_1)\cdot(\m_1-\m_2)\right)^2
		%	\\
		%	&F_0(\s+\m_1+\l_1)F_0(\s+\m_2+\l_1)Z_{j_1}(\s+\m_1+\l_1)Z_{j_2}(\s+\m_2+\l_1)
		%	\end{align} 
		%In the following, $p,q\in{1,...,n}$, $p\neq q$, and mirroring the discussion in the previous section, we put $\n_1=e_p+e_q$ $\n_2=0$ to get on the LHS
		%	\begin{dmath}
		%	t^{\s^2+1+\s\cdot(e_p+e_q)}(1+(e_p+e_q)\cdot\s)^2F_0(\s)F_0(\s+e_p+e_q)
		%	\end{dmath}
		%	for which we need $\m_1=e_p-e_1$, $\m_2=e_q-e_1$ on the RHS, 
		%	\begin{dmath}
		%	t^{\s^2+\12(\m_1^2+\m_2^2)+\l_1\cdot(\m_1+\m_2)+ \l_1^2+\s\cdot(\m_1+\m_2)}\left(\12\m_1^2-\12\m_2^2+(\m_1-\m_2)\cdot(\s+\l_1)\right)^2\\
		%	F_0(\s+\l_1+\m_1)F_0(\s+\l_2+\m_2)
		%	=t^{\s^2+1+\s\cdot(e_p+e_q)}\left((e_p-e_q)\cdot\s\right)^2F_0(\s+e_p)F_0(\s+e_q)
		%	\end{dmath}
		%	In both cases, 
		By repeating the analysis similarly to the previous case,
		the equation, analogous to \eqref{An1loop}, fixing $B_0$ is 
		%\vspace*{-0.2cm}
		\bea\label{1loopfix}
		&&(1+(e_p+e_q)\cdot\s)^2B_0(\s)B_0(\s+e_p+e_q)\notag\\&=&\left((e_p-e_q)\cdot\s\right)^2B_0(\s+e_p)B_0(\s+e_q) \, .
		\eea
		The two cases are distinguished by the corresponding different asymptotic conditions \eqref{asy}.  Indeed, we have
		\bea
		& B_{0}^{[D_{n}]}(\s)=\prod\limits_{i<j}^n {1\over G(1\pm\sigma_i\pm\sigma_j)
			%G(1-\sigma_i-\sigma_j)G(1+\sigma_i-\sigma_j)G(1-\sigma_i+\sigma_j)
		}\notag \\
		& B_{0}^{[B_{n}]}(\s)
		%    &=\prod\limits_{k}{1\over |G(1+\sigma_k)G(1-\sigma_k)|}\prod\limits_{i<j} {1\over |G(1+\sigma_i+\sigma_j)G(1-\sigma_i-\sigma_j)G(1+\sigma_i-\sigma_j)G(1-\sigma_i+\sigma_j)|}\\
		=\left(\prod\limits_{k=1}^n{1\over G(1\pm \sigma_k)}\right)B_{0}^{[D_{n}]}(\s)
		\notag
		%\notag	
		\eea
		Also the recursion relations are the same, upon using the appropriate root systems $R$:
		\vspace*{-0.2cm}
		%	\bea
		%	\label{Drec}&&
		%	k^2 Z_k(\s)=\notag\\&&\sum\limits_{\substack{(\n-\l_1^\vee)^2+j_1+j_2=k\\ \n\in \l_1^\vee+ Q^\vee\, ,j_{1,2}<k}}\frac{\left(j_1-j_2+2\n\cdot\s\right)^2Z_{j_1}(\s+\n)Z_{j_2}(\s-\n)}{\prod\limits_{\substack{\a\in R, n\geq 1, \a\cdot\n=n}}(\a\cdot\s)^{2n}\prod\limits_{k=1}^{n-1}((\a\cdot\s)^2-k^2)^{2n-2k}}\notag \\ && 
		%	-\sum\limits_{\substack{\n^2+i_1+i_2=k 
		%	 \n\in Q^\vee,\, i_{1,2}<k}}\frac{\left(i_1-i_2+2\n\cdot\s\right)^2Z_{i_1}(\s+\n)Z_{i_2}(\s-\n)}{\prod\limits_{\substack{\a\in R, n\geq 1, 
		%	\a\cdot\n=n}}(\a\cdot\s)^{2n}\prod\limits_{k=1}^{n-1}((\a\cdot\s)^2-k^2)^{2n-2k}}
		%	\eea 
		\begin{align*}\label{Drec}
		k^2 Z_k(\s)=\sum\limits_{\substack{(\n-\l_1)^2+j_1+j_2=k\\ \n\in \l_1+ Q\, ,j_{1,2}<k}}Z_{j_1}(\s+\n)Z_{j_2}(\s-\n)
		\\
		\left(j_1-j_2+2\n\cdot\s\right)^2\frac{B_0(\s\pm\n)}{B_0(\s)^2}-\sum\limits_{\substack{\n^2+i_1+i_2=k\\\n\in Q,\, i_{1,2}<k}}Z_{j_1}(\s+\n)
		\\
		\times Z_{j_2}(\s-\n)
		\left(i_1-i_2+2\n\cdot\s\right)^2\frac{B_0(\s\pm\n)}{B_0(\s)^2}
		\end{align*}
		\vspace*{-0.3cm}
		
		\noindent This result is in line with the contour integral formulae for the relevant Nekrasov partition functions. Indeed the poles in the $D_n$ and $B_n$ cases are the same, with different residues.	
		%	note that 
		%	\begin{dmath}
		%	F_{1-loop}^{[D_{n+1}]}(\sigma_1,...,\sigma_n,0)=\prod\limits_{i}^n\frac{1}{G(1+\sigma_i)^2 G(1-\sigma_i)^2} F_{1-loop}^{[D_{n}]}(\sigma_1,...,\sigma_n)\neq \prod\limits_{i}^n\frac{1}{G(1+\sigma_i) G(1-\sigma_i)}F_{1-loop}^{[D_{n}]}(\sigma_1,...,\sigma_n)=F_{1-loop}^{[B_{n}]}(\sigma_1,...,\sigma_n)
		%	\end{dmath}
		%	as the $B_n$ tau function system is not just a folding of $D_{n+1}$. 
		%	
		From the above recursion relation we can compute the 1-instanton terms 
		\bea
		Z_1(\s)=\sum_{k=1}^n 4\sigma_k^2\frac{B_0(\s\pm e_k)}{B_0(\s)^2}=
		\begin{cases*}
			\sum_{k=1}^n \frac{-1}{\prod\limits_{j\neq k}(\sigma_k^2-\sigma_j^2)^2}, & $B_n$\\
			\sum_{k=1}^n \frac{4\sigma_k^2}{\prod\limits_{j\neq k}(\sigma_k^2-\sigma_j^2)^2}, & $D_n$
		\end{cases*}
		\notag
		\eea
		%	\bea
		%	Z_1(\s)&=&\sum_k 4\sigma_k^2\frac{F_0(\s\pm e_i)}{F_0(\s)^2}=\notag\\
		%	&=&\begin{cases*}
		%	\sum_k \frac{1}{\prod\limits_{j\neq k}(\sigma_k^2-\sigma_j^2)^2}, & $B_n$\\
		%	\sum_k \frac{4\sigma_k^2}{\prod\limits_{j\neq k}(\sigma_k^2-\sigma_j^2)^2}, & $D_n$
		%	\end{cases*}
		%	\eea
		and the 2-instantons
		\bea
		Z_2(\s) = \sum\limits_{\a\in Q^\vee, \a^2=2}\frac{-1}{(\a\cdot\s)^2((\a\cdot\s)^2-1)^2\prod\limits_{\b\cdot\a=1}(\b\cdot\s)^2}\notag\\
		+\sum\limits_{k=1}^n \frac{Z_1(\s+e_k)(\sigma_k+\12)^2+Z_1(\s-e_k)(\sigma_k-\12)^2}{\prod\limits_{\b\cdot e_k=\pm 1}(\b\cdot\s)}\notag
		\eea
		and so on. These are easily compared to \cite{Marino:2004cn}.	
		
		We now turn to the analysis of the other classical groups, which is more involved. Indeed, the $\tau$-system reduces to higher order equations which
		produce more complicated recurrence relations to be solved by a case by case analysis.
		We performed explicit checks for $C_{3}$, $C_{4}$ and $C_{5}$ up to two-instantons 
		again in agreement with \cite{Marino:2004cn}.

		%	All the complexity of combinatorial data used in ADHM residue integrals is here reduced to solving linear systems with quadratic constraints on the coroot lattices which comes from equating powers of $t$, $\{t^{\sigma_i}\}_{i=1,..,\text{rk }\mathfrak{g}}$.
		
		
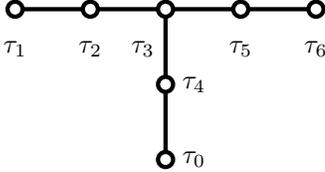
\vspace*{-0.5cm}
		\begin{center}
			\begin{dynkinDiagram}[upside down, ordering=Carter, extended,edge length=1cm, indefinite edge/.style={ultra thick,densely dashed}, edge/.style={ultra thick},o/.style={ultra thick,fill=white,draw=black}, root radius=.1cm,root/.style={ultra thick,fill=white,draw=black}, arrow width = 0.4cm, arrow style={length=5mm, width=5mm,line width = 1pt}]E{oooooo}
				\node[right=.1cm] at (root 0) {$\tau_0$};
				\node[below=.3cm] at (root 1) {$\tau_1$};
				\node[below=.3cm] at (root 2) {$\tau_2$};
				\node[left=.3cm,below=.3cm] at (root 3) {$\tau_3$};
				\node[right=.1cm] at (root 4) {$\tau_4$};
				\node[below=.3cm] at (root 5) {$\tau_5$};
				\node[below=.3cm] at (root 6) {$\tau_6$};
			\end{dynkinDiagram}
		\end{center}
		
		For the exceptional group $E_6$ we obtain the system
		\begin{equation}
		\label{E6}
		\tau_6 D^{4}(\tau_0)=\tau_0 D^{4}(\tau_6) \, .
		\end{equation}
		where we used the notation $D^{2n}:=D^2 \circ D^{2n-2}$.
		The equations which specify $B_0$ can be written as follows. Choose the miniscule weight to be $\l = (0^5,(-{2\over3})^3)$. Let $p_1,...p_5$ be a permutation of $\{1,...,5\}$ and let $\boldsymbol \delta:=((\12)^8)$. Then
		one gets from the lowest order in \eqref{E6}
		%	\bea
		%	\left(1+\sigma_{p_1}+\sigma_{p_2}\right)^2\left(1+\sigma_{p_1}+\sigma_{p_3}\right)^2\left(\sigma_{p_2}-\sigma_{p_3}\right)^2 \times\notag\\
		%	F_0(\s)F_0(\s+e_{p_1}+e_{p_2})F_0(\s+e_{p_1}+e_{p_3})=\notag
		%	\\ \notag =\left((\boldsymbol \delta -e_{p_2}-e_{p_3})\cdot\s\right)^2
		%	\left((\boldsymbol \delta -e_{p_2}-e_{p_3}-e_{p_4}-e_{p_5})\cdot\s\right)^2
		%	\left(\sigma_{p_4}+\sigma_{p_5}\right)^2 \times 
		%	\\F_0(\s+\boldsymbol \delta+\l)F_0(\s+\boldsymbol \delta+\l-e_{p_4}-e_{p_5})F_0(\s+e_{p_1}-\l/2)
		%	\eea
		\bea
		\left(1+\sigma_{p_1}+\sigma_{p_2}\right)^2\left(1+\sigma_{p_1}+\sigma_{p_3}\right)^2\left(\sigma_{p_2}-\sigma_{p_3}\right)^2 \times \notag\\
		B_0(\s)B_0(\s+e_{p_1}+e_{p_2})B_0(\s+e_{p_1}+e_{p_3})= \notag\\
		\left((\boldsymbol \delta -e_{p_2}-e_{p_3})\cdot\s\right)^2
		\left((\boldsymbol \delta -e_{p_2}-e_{p_3}-e_{p_4}-e_{p_5})\cdot\s\right)^2
		\times \notag\\
		\left(\sigma_{p_4}+\sigma_{p_5}\right)^2 
		B_0(\s+\boldsymbol \delta+\l)\times \notag\\
		B_0(\s+\boldsymbol \delta+\l-e_{p_4}-e_{p_5})B_0(\s+e_{p_1}-\l/2) \notag
		\eea
		The solution satisfying the asymptotic behaviour \eqref{asy} is $B^{[E_6]}_0=$ 
		\vspace*{-0.3cm}
		\begin{equation*}
		\prod\limits_{i<j=1}^5 { 1 \over G(1\pm\sigma_i\pm\sigma_j)
			%G(1-\sigma_i-\sigma_j)G(1+\sigma_i-\sigma_j)G(1-\sigma_i+\sigma_j)
		}\prod\limits_{\substack{\varepsilon_i=\pm 1 \\ \prod_{i=1}^8\varepsilon_i = 1, \\ \varepsilon_6=\varepsilon_7=\varepsilon_8}} { 1 \over G(1+{1 \over 2}\sum\limits_{i=1}^8\varepsilon_i \sigma_i) }
		\end{equation*}
		%\vspace*{-0.2cm}
		We also solved the recurrence relation arising from \eqref{E6} up to two-instantons. For one-instanton, our results agree with the ones of \cite{Keller:2011ek}, while the two instantons result
		is a too huge formula to be reported here. 	
		%	we performed an alternative derivation via blow-up equations of \cite{Nakajima:2003pg}.
		%	from
		%	a four-dimensional limit of 5d blow-up equations appearing in \cite{?} finding full agreement. 
		We remark that \eqref{E6} represents a completely novel way of obtaining equivariant volumes of 
		instanton moduli spaces for exceptional groups. 
		
		Unimodular algebras $G_2,F_4,E_8$ have no outer automorphisms and consequently all the $\tau$-functions associated to different nodes are independent. Therefore, the equations on the $\tau$-function associated to the affine node turn out to be more difficult to solve. 	Let us display them for the $G_2$ case.
		%		What remain are the \textit{unimodular} algebras $G_2,F_4$ and $E_8$, in the sense that the determinant of their finite Cartan matrix is unity. Note that although $(F_4^{(1)})^\tau$ can be realized by gluing two "legs" of $E_6^{(1)}$, we do not obtain the exact same equation as $(5)$ with $\tau_6\mapsto\tau_4$. Likewise, the Dynkin diagram $(G_2^{(1)})^\tau$ is obtained by quotienting $D_3^{(1)}$ by a tripotent automorphism, but the system of tau functions is not, not in the sense of \cite{Khastgir1995} as the powers of the tau functions need to be adjusted according to length. On the other hand, $E_8^{(1)}=E_9$ leads to a new equation altogether. In each case, we lack the symmetry of the Coxeter graph which maps the affine node to another, which is reflected in the fact that the center of the corresponding Langlands dual group is trivial, or equivalently, the coroot lattice equals coweight one, or equivalently, the finite Cartan matrix is unimodular. We may attach \eqref{Kiev} to the affine node, but shifts do not exist, therefore these do not fit into the general framework of the previous analysis and should be analyzed differently. 
		%	\\ \\	
		%	For $G_2$ and $F_4$ explicit equations of high order may be gotten in terms of the single tau function associated to the affine node. 
		%	For $E_8$, the fork makes a reduction to a single equation invonving one tau function impossible, so one has to deal with tau functions which are not just B\"acklund transforms of others. 
		
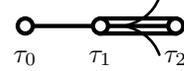
\vspace*{-0.3cm}
		\begin{center}
			\begin{dynkinDiagram}[upside down, reverse arrows, extended,edge length=1cm, indefinite edge/.style={ultra thick,densely dashed}, edge/.style={ultra thick},o/.style={ultra thick,fill=white,draw=black}, root radius=.1cm,root/.style={ultra thick,fill=white,draw=black}, arrow width = 0.4cm, arrow style={length=5mm, width=8mm,line width = 1pt}]G{oo}
				\node[below=.2cm] at (root 0) {$\tau_0$};
				\node[below=.2cm] at (root 1) {$\tau_1$};
				\node[below=.2cm] at (root 2) {$\tau_2$};
			\end{dynkinDiagram}
		\end{center}
		\vspace*{-0.2cm}
		In the normalization where its longest root has length 2, the $G_2$ coroot lattice is the span $Q^\vee=\Z\frac{1}{\sqrt3}(-2,1,1)\oplus\Z\sqrt3(1,-1,0)$. We introduce $\s=(\sigma_1,\sigma_2,\sigma_3)$ but all expressions should be restricted to $\sigma_1+\sigma_2+\sigma_3=0$. By eliminating $\tau_1$ and $\tau_2$, the $\tau$-system reduces to the single equation 
		\vspace*{-0.3cm}
		\begin{equation}\label{G2}
		D^2(\tau_0^{-1}D^4(\tau_0))=3t(D^2(\tau_0))^3\, .
		\end{equation}
		%		where the $3$ comes from the ratio of the squares of the long to short roots. 
		The operator on the l.h.s. of \eqref{G2} turns out to factorize as $D^2(\tau_0^{-1}D^4(\tau_0)) = \tilde{D}^4(\tau_0)\cdot D^2(\tau_0)$, where $\tilde{D}^4(\tau_0)$ is a fourth order operator in $\tau_0$ and its derivatives. The trivial solution of $D^2(\tau_0)=0$ is $\tau_0=a t^b$ which we discard being incompatible with \eqref{Kiev}. In the remainder we insert
		\begin{equation}
		\resizebox{1\hsize}{!}{$\tau_0(\s,\e|t)=\sum\limits_{\n\in Q^\vee}e^{2\pi\I \e\cdot\n}\left(-{t\over 3}\right)^{\12(\s+\n)^2}B\left(\s+\n|-{t\over 3}\right)$}\notag
		\end{equation}
		and obtain, after a rescaling $t\mapsto -3 t$,
		\begin{align}
		&\sum_{\substack{ \{\n_k\}\in Q^\vee\\ \{i_k\}\in{\mathbb N}}} \prod\limits_{k=1}^4 e^{2\pi\I \e\cdot\n_k }t^{\12(\s+\n_k)^2+i_k} B_0(\s+\n_k) Z_{i_k}(\s+\n_k) \notag
		\\
		&\Bigg(\frac{1}{4!}\prod\limits_{k_1<k_2}(\12\n_{k_1}^2+i_{k_1}-\12\n_{k_2}^2-i_{k_2}+(\n_{k_1}-\n_{k_2})\cdot\s)^2\notag
		\\
		&+\frac{9}{4}(\12\n_{1}^2+i_{1}-\12\n_{2}^2-i_{2}+(\n_{1}-\n_{2})\cdot\s)^2\notag
		\\
		&(\12\n_{3}^2+i_{3}-\12\n_{4}^2-i_{4}+(\n_{3}-\n_{4})\cdot\s)^2\Bigg)=0 \, .
		\label{G2explicit}
		\end{align}
		The lowest order terms in \eqref{G2explicit},
		namely the coefficient of $t^{3+\s\cdot({4\over\sqrt{3}},-{2\over\sqrt{3}},-{2\over\sqrt{3}})}$, gives a quartic relation which simplifies to the following quadratic one
		\begin{align*}
		&\left(\frac{2 \sigma_1-\sigma_2-\sigma_3}{\sqrt{3}}+1\right)^2 B_0(\s) B_0\left(\s+\frac{1}{\sqrt3}(2,-1,-1)\right)
		\\
		&=-\left(\frac{\sigma_2-\sigma_3}{\sqrt{3}}\right)^2 
		\left(\frac{\sigma_1+\sigma_2-2 \sigma_3}{\sqrt{3}}\right)^2 
		\left(\frac{\sigma_1-2 \sigma_2+\sigma_3}{\sqrt{3}}\right)^2 
		\\& \times
		\left(\frac{\sigma_1+\sigma_2-2 \sigma_3}{\sqrt{3}}+1\right)^2 
		\left(\frac{\sigma_1-2 \sigma_2+\sigma_3}{\sqrt{3}}+1\right)^2 
		\\ &\times
		B_0\left(\s+\frac{1}{\sqrt3}(1,-2,1)\right)B_0\left(\s+\frac{1}{\sqrt3}(1,1,-2)\right)
		\end{align*}
		By imposing \eqref{asy}, these are solved by $B_0^{[G_2]}(\s)=$
		%\vspace*{-0.2cm}
		\begin{equation*}
		\prod_{i<j}^3{1\over G(1\pm{1\over\sqrt{3}}(\sigma_i-\sigma_j))}\prod_{\substack{ijk\\ cyclic}}^3{1\over G(1\pm{1\over\sqrt{3}}(2\sigma_i-\sigma_j-\sigma_k))}
		\end{equation*}
		%\begin{align*}
		%B_0^{[G_2]}(\s)&=\prod_{i<j}^3{1\over |G(1\pm{1\over\sqrt{3}}(\sigma_i-\sigma_j))|}\\
		%&\times\prod_{\substack{ijk\\ cyclic}}^3{1\over %|G(1\pm{1\over\sqrt{3}}(2\sigma_i-\sigma_j-\sigma_k))|}
		%\end{align*}
		%		or 
		%		\begin{equation}
		%		B_0^{[G_2]}(\s)=B_0^{[A_2]}({1\over\sqrt{3}}\cdot\s)\prod_{\substack{ijk\\ cyclic}}{1\over G(1\pm{1\over\sqrt{3}}(2\sigma_i-\sigma_j-\sigma_k))}
		%		\end{equation}
		%		\\ \\
		The 1-instanton contribution is obtained by considering the coefficient of the next order $t^{3+\s\cdot(\sqrt{3},0,-\sqrt{3})}$ term: all $B_0(\s)$ factors drop out and we obtain just 
		%		\begin{equation}
		%		Z_1(\s)=\frac{486}{(\sigma_1+\sigma_2-2 \sigma_3)^2 (2 \sigma_1-\sigma_2-\sigma_3)^2 (\sigma_1-2 \sigma_2+\sigma_3)^2}
		%		\end{equation}
		%		which is the correct 1-instanton. The next order in $t$ gives us the 2-instanton term in terms of $Z_1(\s)$, $Z_1(\s+\a)$ where $\a^2=2$, and ratios of $B_0$, which agrees with the 4d blowup formula once rewritten as a rational function.
		%		
		%			The 1-instanton part is obtained similarly. In fact, when considering the $t^{3+\s\cdot(\sqrt{3},0,-\sqrt{3})}$ term, all $F_0(\s)$ factors drop out and we obtain just 
		%\vspace*{-0.3cm}
		\begin{equation}\notag
		Z_1(\s)^{[G_2]}\vert_{\sigma_3=-\sigma_1-\sigma_2}=-\frac{2}{3 \sigma_1^2 \sigma_2^2 (\sigma_1+\sigma_2)^2}
		\end{equation}
		in agreement with \cite{Keller:2011ek}. The next order in $t$ gives the 2-instanton term $Z_2(\s)^{[G_2]}\vert_{\sigma_3=-\sigma_1-\sigma_2}=$
		%resized equation:
		\vspace*{-.25cm}
		\begin{equation}\notag
		\resizebox{1\hsize}{!}{$		
			\frac{3 \left(9 \sigma_1^4 \left(6 \sigma_2^2+1\right)+18 \sigma_1^3 \left(6 \sigma_2^3+\sigma_2\right)+3 \sigma_1^2 \left(18 \sigma_2^4+9 \sigma_2^2-2\right)+6 \sigma_1 \sigma_2 \left(3 \sigma_2^2-1\right)+\left(1-3 \sigma_2^2\right)^2\right)}{\sigma_1^2 \left(1-3 \sigma_1^2\right)^2 \sigma_2^2 \left(1-3 \sigma_2^2\right)^2 (\sigma_1+\sigma_2)^2 \left(1-3 (\sigma_1+\sigma_2)^2\right)^2}
			$}\, .
		\end{equation}
		\vspace*{-.5cm}
		
		It should be possible to apply the approach proposed here to general class-$\mathcal{S}$ theories \cite{Gaiotto:2009we} by studying the related isomonodromic deformation problem (for example for linear and circular quivers).
		It would be also interesting to extend the analysis to non-self-dual $\Omega$-background, which should amount to quantum $\tau$-systems, and its lift to five dimensional gauge theories on $\mathbb{R}^4\times S^1$, which should correspond to $q$-difference $\tau$-systems \cite{Bonelli:2017gdk,Bershtein:2018srt,Bonelli:2020dcp}. 
		Finally, it would be great to apply similar ideas to models with less or no supersymmetry, trying to constrain their dynamics by the study of renormalization group equations in presence of surface operators.
		
		\paragraph{\bf Acknowledgements:}
		We would like to thank M. Mari\~no and
		T. Nosaka
		for fruitful discussions. 
		%for useful discussions,  clarifications and a careful reading of the manuscript.
		This research  is  partially supported by the INFN Research Projects GAST and ST$\&$FI, by PRIN "Geometria delle variet\`a algebriche" and  by PRIN
		"Non-perturbative Aspects Of Gauge Theories And Strings".

		\bibliographystyle{unsrtaipauth4-1}
		\bibliography{Biblio}

	\end{document}